# Direct visualization of current induced spin accumulation in topological insulators

Yang Liu[1†], Jean Besbas[1†], Yi Wang[1], Pan He[1], Mengji Chen[1], Dapeng Zhu[1], Yang Wu[1], Jong Min Lee[1], Lan Wang[2], Jisoo Moon[3], Nikesh Koirala[3], Seongshik Oh[3], and Hyunsoo Yang[1*]

[1]Department of Electrical and Computer Engineering, National University of Singapore, 117576 Singapore

[2]RMIT University, School of Science, Physics, Victoria, 3000, Australia

[3]Department of Physics and Astronomy, Rutgers, The State University of New Jersey, Piscataway, New Jersey 08854, USA

[†]These authors contributed equally to this work.

*Corresponding author. E-mail: eleyang@nus.edu.sg

**Charge-to-spin conversion in various material systems is the key for the fundamental understanding of spin-orbitronis as well as the development of efficient means to manipulate the magnetization. We report the direct spatial imaging of current induced spin accumulation at the channel edges of $Bi_2Se_3$ and $BiSbTeSe_2$ topological insulators by a scanning photovoltage microscope at room temperature. The spin polarization is along the out-of-plane direction with opposite signs for the two channel edges. The accumulated spin direction reverses sign upon changing the current direction and the detected spin signal shows a linear dependence on the magnitude of currents, which indicates that our observed phenomena are the current induced effects. The spin Hall angle of $Bi_2Se_3$ and $BiSbTeSe_2$ is**

**determined to be 0.0085 and 0.0616, respectively. We further image the current induced spin accumulation in a Pt heavy metal. Our results open up the possibility of optically detecting the current induced spin accumulations, and thus point towards a better understanding of the interaction between spins and circularly polarized light.**

Charge-to-spin conversion has been considered as one of the core research fields in spintronics[1,2,3,4,5]. Over the past decade, current induced spin accumulation due to the spin Hall effect (SHE) in semiconducting systems has been extensively investigated by means of magneto-optical effects[6,7,8], circularly polarized electroluminescence[9,10] and two-color optical coherence techniques[11]. In metallic systems, electrical detection of current induced spin accumulation has been studied by employing spin valve[12,13,14], spin pumping[15,16] and spin torque ferromagnetic resonance[17,18,19]. While the experimental situation in heavy metals for a direct optical detection of SHE is less clear with two groups suggesting a successful measurement of the SHE in Pt and W using MOKE[20,21] and others concluding that MOKE is not suitable in measuring SHE in metallic systems[22,23].

The three-dimensional topological insulator (TI) is a new phase of the quantum state of materials that possess spin-momentum locked surface states and insulating bulk states[24,25,26,27,28,29]. So far, the surface state related phenomena of TIs responding to light has been extensively explored, including the chemical potential drop assisted photovoltage generation[30,31,32], circularly polarized light induced helicity-dependent current[33] and circular photogalvanic effect in TIs[34]. However, the interaction between the current induced spin accumulation with circularly polarized light is less studied especially involving the direct visualization of current induced spin accumulation in TIs.

The spin-to-charge conversion in TIs has been extensively studied based on ferromagnet (FM)/TI bilayer structures where the FM layer can be used as either a spin injector or detector[35,36,37,38,39,40]. In such FM/TI bilayer systems, the measured spin Hall angle could involve not only intrinsic TI contributions but also the contribution from the FM/TI interface[41,42,43], which can largely modulate the effective spin Hall angle in TIs. Thus it is of great importance to quantitatively characterize the spin-to-charge conversion of TIs without involving a FM layer.

In this work, we use the scanning photovoltage microscope to image current induced spin accumulations near channel edges in $Bi_2Se_3$ and $BiSbTeSe_2$ TIs at room temperature. We further image the current induced spin accumulation in a Pt heavy metal. Our work opens up a possibility of optically detecting the accumulated spins in various metallic and semiconducting materials, and helps to extract spin-related parameters such as the spin Hall angle and spin lifetime, propelling a better understanding of the interactions between spins and light in various materials systems.

## Results

**Principle of helicity dependent photovoltages**

Figure 1a and 1b show the experimental geometry of scanning photovoltage microscope. The laser is normally incident on the device, while a direct bias current is applied along the $y$ axis (Fig. 1a). To investigate the effects of the laser helicity on the photovoltage generation, a photoelastic modulator acting as a rotating quarter waveplate at a frequency of $f_{\text{PEM}} \approx 50$ kHz is used to modulate the helicity of light. The magnitude of the helicity dependent photovoltages is proportional to the difference between the photovoltages generated by left circularly polarized (LCP) and right circularly polarized light (RCP). The helicity dependent photovoltage (HDP) is thus defined as $V_{\pm} = V_{RCP} - V_{LCP}$. By employing PEM, a helicity independent background signal due to the thermoelectric effect in TIs can be substantially suppressed in our HDP measurements.

In general, the helicity dependent photovoltages appear in materials with spin-split band structures. When shining circularly polarized lights on the materials with strong spin orbit coupling normally, LCP and RCP light generate electrons with opposite spin direction. If no bias current is applied across the device, the absorption of LCP and RCP light is equal. Thus, the induced HDP is zero ($V_{\pm} = V_{RCP} - V_{LCP} = 0$). When applying bias currents, the current induced out-of-plane spins accumulate near two opposing channel edges (Fig. 1a). As the amount of absorption of the LCP and RCP light changes depending on local spin accumulation due to current induced effects, magnetic circular dichroism, which is a difference between the absorption of RCP and LCP light, is induced at the edges of the device channel. Therefore, the sign and magnitude of induced HDP provide information of the local spin polarization distribution.

**Scanning helicity dependent photovoltage data in $Bi_2Se_3$ and Pt**

Figure 2a-e show the scanning HDP data in a $Bi_2Se_3$ channel (labeled as BS1) with different bias currents. The boundaries of the device channel are indicated by black dashed lines while black arrows show the direction of the applied bias current with the current density on the order of $10^6$ A cm$^{-2}$. There is a negligible signal in the absence of bias currents in Fig. 2c. On the other hand, stronger signals are detected at the two edges of the device channel when the magnitude of bias currents increases. The signs of signals are opposite at two channel edges, and reversing the bias current direction causes the HDP to switch its sign. Considering the geometry of the electrical connections with the spin accumulation signals at edges of the device, we establish that the sign of the spin Hall angle $\theta_{sh}$ is positive in $Bi_2Se_3$, in good agreement with the previous results in TI/FM structures[35,36,37,38].

In order to check whether our method can be applied to other material systems, HDP measurements are performed on a heavy metal Pt, which is known to have a strong current induced spin accumulation with a positive sign of $\theta_{sh}$[17,42,44]. Fig. 2f-j show similar results as Bi$_2$Se$_3$, but with larger bias currents required to obtain similar HDP signal intensities due to overwhelming background conduction electrons in a metal. As a control experiment, we have performed measurements on Cu (Supplementary Note 2), which is known to have small spin orbit coupling. There is no observable signal in Cu either without or with bias current (Supplementary Fig. 4). We can also rule out the contribution from current induced heating as reversing the current direction causes the HDP to switch its sign, indicating that our observed phenomena in Bi$_2$Se$_3$ and Pt are indeed current induced spin accumulation due to sizable spin orbit interaction.

**Bias current and magnetic field dependence in Bi$_2$Se$_3$**

We have further fixed the laser near one edge on another Bi$_2$Se$_3$ device (labeled as BS2) and measured the longitudinal HDP voltage ($V_L$) while sweeping the bias current (Fig. 3a). As shown in Fig. 3b, the linear relationship of $V_L$ with respect to the bias current suggests that the observed phenomenon in Bi$_2$Se$_3$ is indeed a current induced effect.

We have then performed Hanle measurements on BS2 to evaluate the spin lifetime. An in-plane external magnetic field ($B_{ext}$) was applied perpendicular to the bias current direction to induce spin precession (Fig. 3c). Figure 3d shows the longitudinal HDP data obtained by locating the laser at one edge of the channel and sweeping the magnitude of the $B_{ext}$. The data are fitted to $A_1/[1+(\Omega\tau_s)^2]$, where $A_1$ is the peak of HDP, $\Omega = g\mu_B B_{ext}/\hbar$, $g$ is g-factor, $\mu_B$ is Bohr magneton and $\hbar$ is reduced Plank constant[6]. The spin lifetime ($\tau_s$) of $Bi_2Se_3$ is evaluated to be ~ 3.3 ± 0.13 ps (Supplementary Note 3), which is similar to the reported values in the previous studies[30,45,46,47].

**Extraction of spin Hall angle in $Bi_2Se_3$ and the role of bulk spin Hall effect**

We next evaluate the spin Hall angle $\theta_{sh}$ in BS2, where the HDP is detected along the transverse direction ($V_T$, Fig. 3e) with respect to the bias current[48]. In our spin Hall angle evaluation method, circularly polarized light incident on the Hall-bar device perpendicular to the sample plane excites out-of-plane spin-polarized carriers, which generate a spin dependent transverse voltage $V_T$ due to ISHE[48,49]. $V_T$ can be written as $V_T = \theta_{sh}\rho_N w j_{//} P$, where $\theta_{sh}$ is spin Hall angle, $\rho_N$ is the resistivity of the sample, $w$ is the channel width, $j_{//}$ is the bias current density and $P$ is the carrier spin polarization. We obtain the spin Hall angle $\theta_{sh}$ ~ 0.0085 ± 0.0016 (Supplementary Note 4), in line with previous findings of $\theta_{sh}$ in $Bi_2Se_3$[37,38]. Recent experiments using various transport methods demonstrated a wide range of the charge-to-spin conversion efficiency in $Bi_2Se_3$ from 0.009 to 3.5[35,36,37,38,50,51]. The $\theta_{sh}$ measured in our work is similar to the value obtained from spin pumping, indicating that the spin currents generated from the bulk contributes significantly to our observed signal[37,38].

We further discuss the contribution of topological surface states (TSS) and bulk states (BS) to the out-of-plane edge spin accumulation in $Bi_2Se_3$. We can rule out the two-dimensional electron gas (2DEG) states as the major contribution for spin accumulation since the 2DEG provides the opposite spin polarity compared to the observed spin accumulation direction in Fig. 2a-e[52,53,54]. By employing the multi-channel model as reported previously[55,56], the ratio of the spin current flowing in TSS and BS is $I_{s\text{-}TSS}:I_{s\text{-}BS} = 1:2.06$ considering the thickness of TSS to be ~ 1 nm and BS to be ~ 1.9 nm (Supplementary Note 5)[52,56,57], indicating a considerable contribution of the bulk spin Hall effect in 9 QL $Bi_2Se_3$. We perform the thickness dependence study on $Bi_2Se_3$. The HDP magnitude increases significantly when $t_{BiSe} < 9$ QL as shown in Supplementary Fig. 5, indicating a considerable bulk spin Hall effect in 9 QL $Bi_2Se_3$ at room temperature (Supplementary Note 5).

**Helicity dependent photovoltages in BiSbTeSe$_2$**

We have performed similar measurements on BiSbTeSe$_2$ (labeled as BSTS1), which was reported to have a larger spin-charge conversion efficiency. As shown in Fig. 4a and 4c, the spin polarization has opposite signs for the two edges and changes the sign on reversing the current direction, similar to the data from $Bi_2Se_3$. In contrast, there is a negligible signal in the absence of bias currents in Fig. 4b. Under the presence of bias currents, the $V_L$ is proportional to the bias current (Fig. 4d), indicating the current induced effect. Furthermore, we have performed the magnetic field dependence study (Fig. 4e) on BiSbTeSe$_2$ by fixing the laser near one edge of another device (BSTS2). The data are fitted to the Hanle equation and the spin lifetime is extracted to be ~ 18.6 ± 1.5 ps (Supplementary Note 3). We then measure the $V_T$ in BSTS2 (Fig. 4f), and the spin Hall angle in BSTS2 is evaluated to be ~ 0.0616 ± 0.0101 (Supplementary Note 4). We further calculate the spin current flowing in TSS and BS by employing the multi-channel model. The ratio of the spin current flowing in TSS to BS spin current is $I_{s\text{-}TSS}$:$I_{s\text{-}BS}$ = 1:1.62, indicating a sizable bulk contribution at room temperature. In line with the above results, the Fermi level of BSTS2 is estimated to be located inside the bulk conduction band (Supplementary Note 6)[58].

In summary, we have shown that the current induced spin accumulation not only in topological insulators but also in heavy metals can be detected and imaged by a helicity dependent scanning photovoltage setup at room temperature. The required sample structure is free from a ferromagnet, thus eliminating the interface transparency issue and current shunting problem in a bilayer structure, which typically exist in other electrical characterization methods. The accumulation spin directions at channel edges, the spin lifetime, and the spin Hall angle can be estimated using this technique, which propels the characterization of novel materials and helps for a better understanding of the interaction among charge, spin, and light.

*Note added:* while preparing the revised manuscript, we became aware of a similar work by other group[59].

**Methods**

**Sample and device fabrication.** We employ a previously reported two-step procedure to grow 10 quintuple layer (QL, 1 QL ≈ 1 nm) $Bi_2Se_3$ on $Al_2O_3$ (0001) substrates[56,60]. The high quality molecular beam epitaxy (MBE) grown $Bi_2Se_3$ thin film was then patterned into Hall bar devices (Fig. 1b) by the following processes. First, the Se layer was decapped by annealing in the vacuum chamber. Then, we etched the film by Ar-ion milling for 2 second and a MgO (1 nm)/$SiO_2$ (3 nm) capping layer is sputtered onto the $Bi_2Se_3$ film at room temperature with a base pressure of $3\times10^{-9}$ Torr. The final thickness of the $Bi_2Se_3$ film is 9 nm. Then the film was patterned into a Hall bar by photolithography and Ar ion milling. The dimension of the Hall bar device is 450 μm in the longitudinal direction, 100 μm in the transverse direction and the bar width is 10 μm. In the next step, a top electrode of Ta (5 nm)/Cu (100 nm)/Ru (5 nm) was deposited.

High quality $BiSbTeSe_2$ (BSTS) single crystals were grown by the modified Bridgeman technique. We exfoliated ~ 100–150-nm-thick BSTS nanoflakes onto the n-doped Si wafers with a 300-nm-thick thermal $SiO_2$ layer. An MgO (1 nm)/$SiO_2$ (3 nm) capping layer was sputtered. The Hall bar devices (Supplementary Fig. 2a) was patterned by electron beam lithography (EBL) followed by Ar-ion milling. The device dimension is 30 μm in the longitudinal, 10 μm in the transverse direction and the bar width is 4 μm. Then a top electrode of Ta (5 nm)/Cu (100 nm)/Ru (5 nm) was deposited.

The 6-nm-thick Pt thin film was deposited by magnetron sputtering. An MgO (1 nm)/$SiO_2$ (3nm) capping layer was then sputtered on top of Pt thin film. The Hall bar device was patterned by standard photolithography and Ar-ion milling. The device dimension is 450 μm in the longitudinal, 100 μm in the transverse direction and the bar width is 10 μm. Finally, the top electrode of Ta (5 nm)/Cu (100 nm)/Ru (5 nm) was deposited. We repeated the same fabrication process to fabricate a 6-nm-thick copper Hall bar device for the control experiment.

**Measurements.** In all experiments, laser light (wavelength $\lambda = 650$ nm, power of pump laser beam $I = 4.62$ mW) was focused at normal incidence onto bare $Bi_2Se_3$, $BiSbTeSe_2$ or Pt using a 100x microscope objective lens (Supplementary Fig. 1). The laser spot size is ~ 1 μm and the spatial resolution is set by the size of laser spot. The laser beam was split into two after reflecting from the sample; one was collected by a CMOS detector to display the position of the laser spot on the device, while the other one was detected by a photodetector to measure the reflectivity of the device. Two-dimensional image was achieved by scanning the device with the laser spot while moving the device with a piezo stage.

To measure the helicity dependent photovoltage (HDP), the light circular polarization was modulated while the light intensity was constant. Light circular polarization was modulated using a photoelastic modulator (PEM). In this configuration, the PEM acts as a rotating quarterwave plate and modulates the polarization of the laser at a fixed frequency of 50 kHz and without affecting the laser intensity. The angle of light linear polarization before the PEM was set to 45º with respect to the axis of the PEM. The HDP between the device electrodes was acquired by a lock-in amplifier.

**Data availability.** The data that support the plots within this paper and other findings of this study are available from the corresponding author upon reasonable request.

**Figure 1 | Schematics spin-dependent photovoltage generation and experimental setup. a,** Schematic of the device structure with bias current $J_c$ applied along the $y$ direction. The laser is focused on the device with normal incidence. **b**, Optical image of the device with a representation of the electrical connections. Scale bar is 50 µm. The scanning area is indicated by a yellow box. The induced photovoltage is acquired by a lock-in amplifier.

**Figure 2 | Current induced spin accumulations in Bi$_2$Se$_3$ and Pt. a-e,** Spatial two-dimensional HDP maps in BS1 under bias currents of 1 (**a**), 0.8 (**b**), 0 (**c**), -0.8 (**d**), and -1 mA (**e**). **f-j**, Spatial two-dimensional HDP maps in Pt under bias currents of 6 (**f**), 3 (**g**), 0 (**h**), -3 (**i**), and -6 mA (**j**). Black dashed lines indicate the edges of the device. Black arrows show the applied bias current direction.

**Figure 3 | Current dependence of the spin accumulations in Bi$_2$Se$_3$. a,** Experimental geometry of longitudinal photovoltage ($V_L$) measurement. **b**, Bias current dependence of $V_L$ by fixing the laser spot at one edge of the device in BS2. **c**, Experimental geometry of Hanle measurement. **d**, Measurement of HDP at one edge as a function of in-plane $B_{ext}$ in BS2. **e**, Experimental geometry of transverse photovoltage ($V_T$) measurement. **f**, Bias current dependence of $V_T$ by fixing the laser spot at the center of the Hall cross in BS2. Solid lines are fits.

**Figure 4 | Current induced spin accumulations in BiSbTeSe$_2$. a-c**, Spatial two-dimensional HDP map in BSTS1 under bias currents of 0.3 (**a**), 0 (**b**), -0.3 mA (**c**). Black dashed lines represent the boundaries of the device. Black arrows show the direction of applied bias current. **d**, Bias current dependence of $V_L$ by fixing the laser spot at one edge of the device in BSTS2. **e**, Measurement of HDP at one edge as a function of $B_{ext}$ in BSTS2. **f**, Bias current dependence of $V_T$ by fixing the laser spot at the center of the Hall cross in BSTS2. Solid lines are fits.

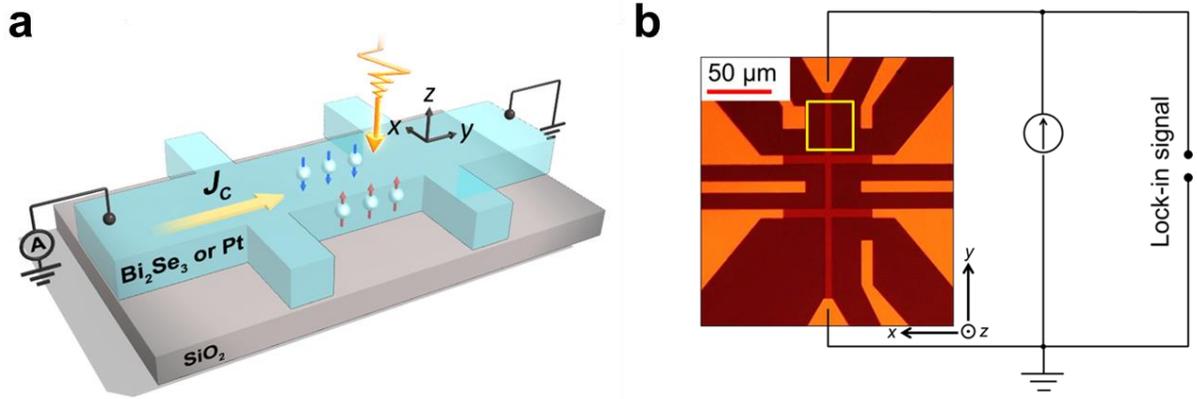

**Figure 1**

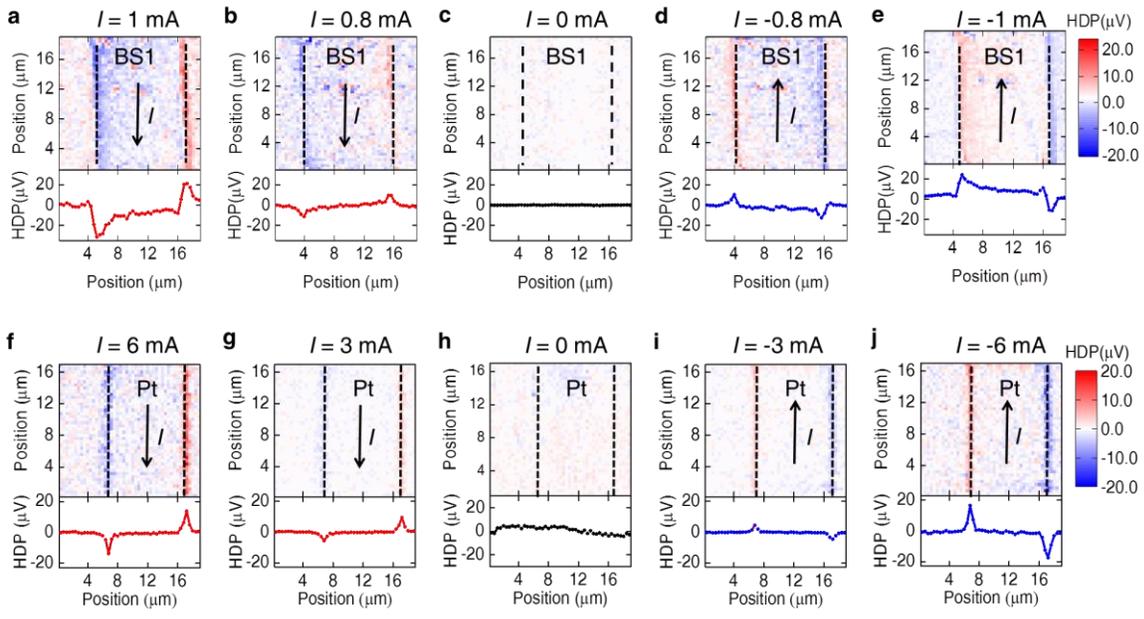

**Figure 2**

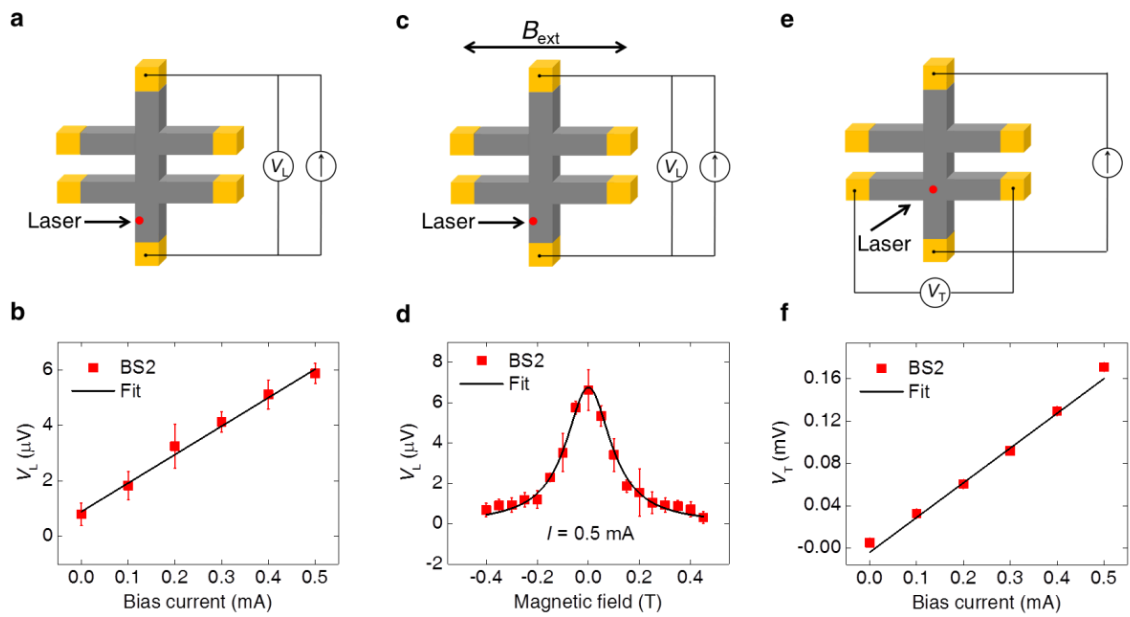

**Figure 3**

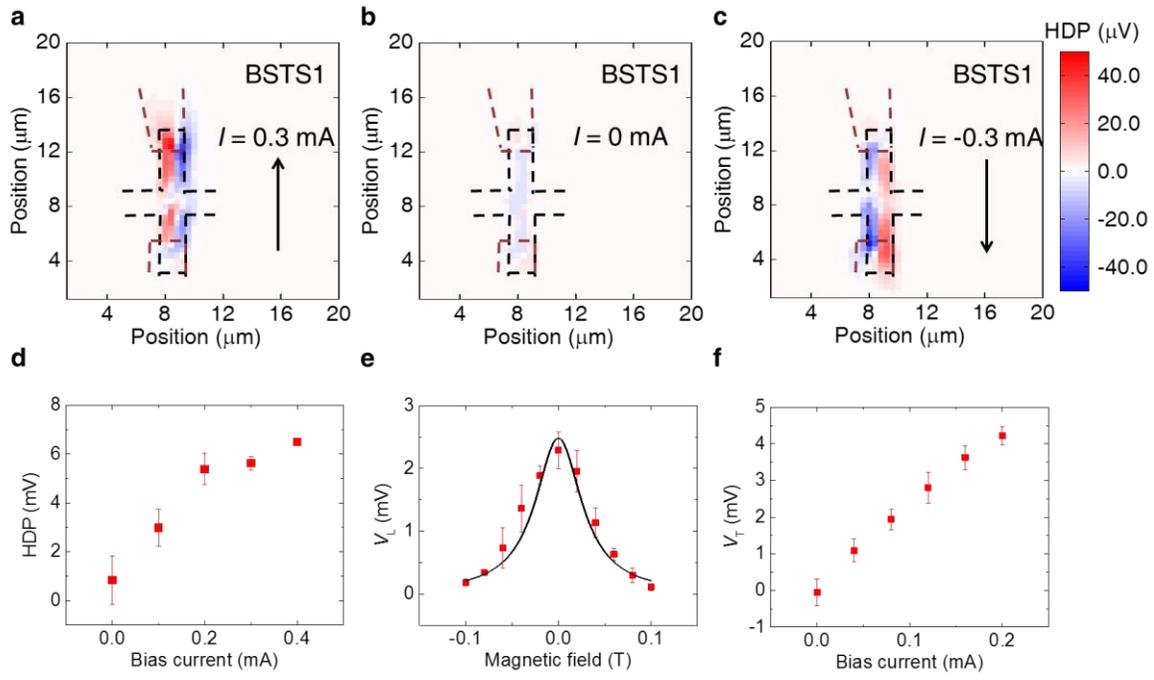

**Figure 4**